\documentclass[11pt, final]{article}

\usepackage{setspace}
\usepackage[all]{nowidow}
\usepackage[english]{babel}
\usepackage{microtype}

\usepackage{amsmath}
\usepackage{amssymb}
\usepackage{bm}
\usepackage{physics}

\usepackage{cite}
\usepackage{authblk}
\usepackage{hyperref}

\usepackage{tikz}
\usepackage{tabularx}
\usepackage{array}
\usepackage{multicol}
\usepackage[font=footnotesize]{caption}
\newcolumntype{C}{>{$}c<{$}}                
\usetikzlibrary{arrows, positioning}

\usepackage{changes}
\usepackage{ifdraft}
\ifdraft{\usepackage[left=5mm, right=45mm, marginparwidth=40mm]{geometry}}{\usepackage{geometry}}

\newcommand{\mb}{\mathbb}
\newcommand{\mc}{\mathcal}

\newcommand{\normord}[1]{\vcentcolon\mathrel{#1}\vcentcolon}    
\providecommand{\vcentcolon}{\mathrel{\mathop{:}}}
\numberwithin{equation}{section}

\begin{document}

\title{Superconformal symmetry in a class of Schellekens theories}
\author{Harry Fosbinder-Elkins\thanks{harryfe@uchicago.edu} }
\author{Jeffrey A. Harvey\thanks{j-harvey@uchicago.edu}}
\affil{Leinweber Institute for Theoretical Physics, \\ Enrico Fermi Institute and Department of Physics \\ University of Chicago \\ Chicago, IL 60637, U.S.A.}

\maketitle

\abstract{In \cite{moore_beauty_2023}, Moore and Singh used the theory of orbifold vertex operator algebras to explicitly construct an $\mc N = 1$ supercurrent in the Beauty and the Beast module of \cite{dixon_beauty_1988}. Using their techniques, we show that $\mc N = 1$ superconformal symmetry exists in all VOAs constructed as spin lifts of canonical $\mathbb{Z}_2$ orbifolds of Niemeier lattice VOAs.}
\clearpage

\section{Introduction}
Holomorphic conformal field theories at central charge $c=24$ form a distinguished class of vertex operator algebras (VOAs) with modular invariant partition functions, including the famous moonshine module constructed by Frenkel, Lepowsky, and Meurman \cite{frenkel_vertex_1988}. Moreover, these VOAs now enjoy a complete classification. This classification was originally conjectured by Schellekens \cite{schellekens_meromorphic_1993} who classified the possibilities for the dimension-one part, $V_1$, of a $c=24$ holomorphic VOA and showed that there were $70$ possible choices. In addition, he conjectured that each choice of $V_1$ leads to a unique VOA. Following a number of partial results, this classification has now been proven. See \cite{Hohn:2017dsm, MR4513145, MR4200469, Hohn:2023auw, MR4810075} for an overview of the classification in terms of generalized deep holes of the Leech lattice. From a conformal field theory point of view,  the classification shows that all Schellekens theories can be constructed as orbifolds of the lattice VOA associated to the Leech lattice $\Lambda_L$ and classifies the orbifolds in terms of a shift by a deep hole of the Leech lattice and the action of an element of the Conway group $Co_0={\rm Aut}(\Lambda_L)$.

Each even unimodular lattice in 24 dimensions (a Niemeier lattice) gives rise to a bosonic $c=24$ vertex operator algebra; these are known as Niemeier theories. Each such lattice has a canonical $\mb Z_2$ symmetry which inverts the sign of all lattice vectors. In this language, the moonshine module is a $\mb Z_2$ orbifold by this canonical $\mb Z_2$ of the Niemeier theory associated to the Leech lattice.  

In 1988, Dixon, Ginsparg, and the second author \cite{dixon_beauty_1988} argued that the moonshine module as well as the $\mb Z_2$ orbifolds of the Niemeier theories admit an $\mathcal{N}=1$ superconformal extension which is often called the Beauty and the Beast model.
\footnote{This was described somewhat vaguely as a $\mb Z_2$ extension in \cite{dixon_beauty_1988}. In modern terminology, it involves a spin lift \cite{duncan_super-moonshine_2006,kapustin_fermionic_2017}.}
This proposal provided early evidence that supersymmetry might be realized in (extensions of) $c=24$ chiral theories. However, while superconformal symmetry was strongly suggested in these models, the actual chiral supercurrent was not constructed explicitly in that work. The existence of the chiral supercurrent in the Beauty and the Beast model was first established in \cite{Gaiotto:2018ypj} using a specific code construction. 

The supercurrent in the Beauty and the Beast model was identified using more general VOA techniques by Moore and Singh in \cite{moore_beauty_2023}. The authors reduced the problem of constructing an $\mathcal{N}=1$ supercurrent within the Beauty and the Beast model to that of finding a certain type of sublattice within the Leech lattice; a problem which they were subsequently able to solve. This result served to ``apprehend'' the missing supercurrent in the Beauty and the Beast theory.

In this paper, we show that the phenomenon uncovered by Moore and Singh is in fact completely general across the Niemeier landscape: \emph{all} $\mathbb{Z}_2$-twisted Niemeier lattice VOAs admit an $\mathcal{N}=1$ superconformal structure.  In other words, starting from any Niemeier lattice $\Lambda$, one can orbifold the associated $c=24$ VOA by the $\mathbb{Z}_2$ reflection and then perform the canonical spin lift of the orbifold theory, and the resulting chiral CFT will contain a supercurrent. We construct such supercurrents in each case by identifying a suitable \emph{superconformal sublattice} $\Lambda_{SC} < \Lambda$.  Roughly, $\Lambda_{SC}$ is a rank $2^{12}$ sublattice of $\Lambda$ with minimum squared norm greater than 4. In the lattice VOA associated to the Niemeier lattice, the superconformal sublattice allows us to produce a weight $\frac{3}{2}$ field with the operator product expansions (OPEs) of a supercurrent, thus generating an $\mathcal{N}=1$ super-Virasoro algebra. Through this construction, we show that every Niemeier-based $\mathbb{Z}_2$ orbifold (including all 24 cases: the 23 Niemeier lattices with roots and the Leech lattice case) possesses an $\mathcal{N}=1$ supercurrent, obtained in a uniform and explicit way.

Previously, this construction was known to yield a supersymmetric theory only for the Leech lattice (the Beauty and the Beast theory), the $A_1^{24}$ Niemeier lattice, and the $A_2^{12}$ Niemeier lattice \cite{dixon_beauty_1988,Gaiotto:2018ypj,benjamin_extremal_2015,harrison_extremal_2016}. Interestingly, in the latter two cases, the resultant theory possesses not just $\mc N = 1$ but higher SUSY, and is a so-called ``extremal SCFT,'' meaning that the theory posesses the largest gap in conformal weights allowed by conformal symmetry and SUSY. Such theories are conjecturally the holographic duals of pure supergravity \cite{witten_three-dimensional_2007}.

We now summarize the structure of the paper and the main technical ingredients of our work. In section 2 we review the construction of Niemeier lattice VOAs and their $\mathbb{Z}_2$ reflection orbifolds, including the definition of the spin-lifted theory and the general properties of twisted sectors. Section 3 is devoted to the extension of the Moore–Singh construction to arbitrary Niemeier lattices.  There we introduce the notion of a superconformal sublattice and show how to explicitly build an $\mathcal{N}=1$ supercurrent in each $\mathbb{Z}_2$-twisted Niemeier VOA. Section 4 discusses the existence of superconformal sublattices in each of the Niemeier lattices and describes our approach to efficiently finding these structures. Finally, section 5 offers some concluding remarks, including implications for the broader classification of $c=24$ SCFTs and open questions for future work. Our main result – the constructive proof that all Niemeier orbifold theories admit $\mathcal{N}=1$ supercurrents – thus completes the supersymmetric extension of this rich class of $c=24$ VOAs.

\section{Niemeier theories and $\mb Z_2$ orbifolds}
To any even, self-dual lattice of rank $r$ one may associate a bosonic, holomorphic CFT with central charge $r$ in a canonical way \cite{frenkel_vertex_1988,dolan_conformal_1990,dolan_conformal_1996}. Any such theory enjoys a $\mb Z_2$ automorphism due to the reflection symmetry of the lattice; we may orbifold by this $\mb Z_2$ to obtain a new theory with a $\mb Z_2$ twisted sector. Moreover, the orbifold theory may be enlarged in a canonical way to form the so-called ``spin lift.'' To be precise, we start with the bosonic (unorbifolded) theory, stack the theory with a 2D TQFT known as the Arf invariant in such a way that the $\mb Z_2$ automorphism of the bosonic theory is coupled to the Arf theory, and then orbifold by the ``diagonal'' $\mb Z_2$ symmetry of both theories \cite{kapustin_fermionic_2017}. If the lattice in question is the Leech lattice (the unique even, self-dual lattice in 24 dimensions with no vectors of squared-length 2), the theory resulting from this procedure is the ``Beauty and the Beast'' CFT \cite{dixon_beauty_1988,moore_beauty_2023}. However, there are 23 other even, self-dual lattices in 24 dimensions, known as the Niemeier lattices, and the procedure may be carried out beginning with any of them. These canonical $\mb Z_2$ orbifolds of the other Niemeier theories were studied in the physics literature before the conjectural classification of Schellekens\cite{Dolan:1989kf}.\footnote{In Table \ref{orbtab} we have corrected one of the entries in table 1 of \cite{Dolan:1989kf}.} These orbifolds lead either back to a Niemeier theory or to Schellekens theories shown in Table \ref{orbtab}. In this section, we collect some useful properties of the theories resulting from this procedure.

\begin{table}[h]
  \centering
  \begin{tabular}{|l|l|l|}
    \hline
    \textbf{Schellekens \#} & \textbf{Affine Lie Algebra} & \textbf{Niemeier lattice} \\
    \hline
    $61$ & $D_{8,1}^3$ & $E_8^3$ \\
    $42$ & $D_{4,1}^6$ & $D_{16} E_8$  \\
    $15$ & $A_1^{24}$ & $D_8^3$ \\
    $ $ & $ \emptyset $ & $\Lambda$ \\
    $66$ & $ D_{12,1}^2$ & $D_{24}$ \\
    $54$ & $ D_{6,1}^4 $ & $D_{12}^2$ \\
    $30$ & $ A_{3,1}^8 $ & $D_6^4$  \\
    $49$ & $ D_{5,1}^2 A_{7,1}^2 $ & $D_{10} E_7^2$  \\
   \hline
   $5$ & $A_{1,2}^{16}$ & $A_3^8$ \\
    $25$ & $D_{4,2}^2 C_{2,1}^4$ & $D_5^2 A_7^2$  \\
    $57$ & $B_{12,2}$ & $A_{24}$ \\
    $50$ & $D_{9,2} A_{7,1}$  & $A_{17} E_7$ \\
    $47$ & $D_{8,2} B_{4,1}^2$  & $A_{15} D_9$ \\
    $41$ & $B_{6,2}^2$ & $A_{12}^2$ \\
    $39$ & $D_{6,2}^2 C_{4,1} B_{3,1}^2$  & $A_{11} D_7 E_6$ \\
    $31$ & $D_{5,2}^2 A_{3,1}^2$ & $A_9^2 D_6$  \\
    $29$ & $B_{4,2}^3$  &  $A_8^3$ \\
    $23$ & $B_{3,2}^4$  & $A_6^4$ \\
    $16$ & $A_{3,2}^4 A_{1,1}^4$ & $A_5^4 D_4$ \\
    $12$ & $C_{2,2}^6$ & $A_4^6$  \\
    $2$ & $A_{1,4}^{12}$  &  $A_2^{12} $\\
    $38$ & $C_{4,1}^4$  & $E_6^4$ \\
    \hline
  \end{tabular}
  \caption{Schellekens theories specified by their affine Lie algebra arising as canonical $\mb Z_2$ orbifolds of Niemeier theories (third column). Those above the horizontal line are equivalent to Niemeier theories.}
  \label{orbtab}
\end{table}

\subsection{Untwisted lattice VOAs}
Let us begin by describing the untwisted sector of the theory, following \cite[section 3]{dolan_conformal_1990}. Given an even, self-dual, rank $r$ Euclidean lattice $\Lambda$ (with $r \equiv 0 \mod 8$), define the holomorphic fields $X^i(z)$ whose target space is the torus $\mb R^r/\Lambda$. The fields have the mode expansion\footnote{
N.B. we use $i$, $j$, $k$ for spacetime indices and $n$, $m$, $l$, $s$, $t$ for summation variables throughout this work.
}
\begin{equation}
    X^i(z) = q^i - i p^i \ln(z) + i \sum_{n \neq 0} \frac{a^i_n}{n} z^{-n}
\end{equation}
where $q^i$ and $p^i \equiv a^i_0$ satisfy
\begin{equation}
    \comm{q^i}{p^j} = i \delta^{i j}
\end{equation}
and $a^i_n$ are oscillators satisfying
\begin{equation} \label{eq:Niemeier:untwistedOscillators}
    \comm{a^i_n}{a^j_m} = n \delta^{i j} \delta_{n, -m}
\end{equation}
with $\qty(a_n^i)^\dag = a_{-n}^i$. Introducing, for each $\lambda \in \Lambda$, the highest-weight state
\begin{equation}
    \ket{\lambda}, \qq{} a^i_n \ket{\lambda} = 0\ \forall\ n > 0, \qq{} p^i \ket{\lambda} = \lambda^i, \qq{} e^{i \mu \cdot q} \ket{\lambda} = \ket{\lambda + \mu},
\end{equation}
the Hilbert space $\mc H_\Lambda$ of the theory is spanned by the Fock spaces
\begin{equation} \label{eq:Niemeier:genericState}
    \psi = \qty(\prod_{a = 1}^M a_{-n_a}^{i_a}) \ket{\lambda}
\end{equation}
where $n_a$ are nonnegative integers which may coincide for different values of $a$. The holomorphic stress tensor is, as usual,
\begin{equation}
    T(z) = -\frac{1}{2} \normord{\partial X(z) \cdot \partial X(z)}
\end{equation}
with Laurent expansion
\begin{equation} \label{eq:Niemeier:untwistedStressTensor}
    T(z) = \sum_{n \in \mb Z} L_n z^{-n - 2}, \qq{} L_n = \frac{1}{2} \qty(\sum_{m \in \mb Z} \normord{a_m \cdot a_{n - m}} + \delta_{n, 0} p^2),
\end{equation}
where the normal ordering $\normord{}$ is defined by
\begin{equation}
    \normord{q^i p^i} = q^i p^i \qq{and} \normord{a^i_n a^i_{-n}} = a^i_{-n} a^i_n \ \ \qty(n > 0).
\end{equation}
The modes satisfy the Virasoro algebra
\begin{equation}\label{eq:Niemeier:VirasoroAlg}
    \comm{L_n}{L_m} = (n - m) L_{n + m} + \frac{r}{12} n (n^2 - 1) \delta_{n, -m}
\end{equation}
and the generic state $\psi$ written in (\ref{eq:Niemeier:genericState}) has conformal dimension
\begin{equation} \label{eq:Niemeier:untwistedConformalDimension}
    L_0 \psi = \qty(\frac{p^2}{2} + \sum_{a = 1}^M n_a) \psi.
\end{equation}
Under the state-operator correspondence, the vertex operator associated to the state $\psi$ of (\ref{eq:Niemeier:genericState}) is
\begin{equation}
    V(\psi, z) = \normord{\qty(\prod_{a = 1}^M \frac{i}{(n_a - 1)!} \partial^{n_a} X^{i_a})} \sigma_\lambda
\end{equation}
where $\sigma_\lambda$ are cocycle operators satisfying
\begin{equation}
    \hat{\sigma}_\lambda \hat{\sigma}_\mu = (-1)^{\lambda \cdot \mu} \hat{\sigma}_\mu \hat{\sigma}_\lambda \ \ \qty(\hat{\sigma}_\lambda = e^{i \lambda \cdot q} \sigma_\lambda),
\end{equation}
which we introduce to ensure mutual locality, i.e. that
\begin{equation} \label{eq:Niemeier:mutualLoc}
    V(\psi, z) V(\phi, w) = V(\phi, w) V(\psi, z)
\end{equation}
under analytic continuation. The state associated to the stress tensor under this correspondence is known as the conformal state:
\begin{equation} \label{eq:Niemeier:conformalState}
    T(z) = V(\psi_L, z),
\end{equation}
and
\begin{equation} \label{eq:Niemeier:conformalStateExplicit}
    \psi_L = \frac{1}{2} \sum_i a^i_{-1} a^i_{-1} \ket{0}.
\end{equation}
The defining property of the vertex operators is that
\begin{equation} \label{eq:Niemeier:definingProp}
    V(\psi, z) \ket{0} = e^{z L_{-1}} \psi,
\end{equation}
or equivalently
\begin{equation} \label{eq:Niemeier:definingPropBrokenOut}
    \comm{L_{-1}}{V(\psi, z)} = \dv{V(\psi, z)}{z} \qq{and} \lim_{z \to 0} V(\psi, z) \ket{0} = \psi.
\end{equation}
In fact, $V(\psi, z)$ is the unique operator satisfying (\ref{eq:Niemeier:definingProp}) and mutual locality with the full system of vertex operators. Later on, we will need the fact that if $\psi$ is a conformal primary state,
\begin{gather}
    \comm{L_0}{V(\psi, z)} = \qty(z \dv{}{z} + h_\psi) V(\psi, z), \label{eq:Niemeier:L0WithV} \\
    \comm{L_m}{V_n(\psi)} = \qty((h_\psi - 1) m - n) V_{m + n}(\psi), \label{eq:Niemeier:LmWithVn}
\end{gather}
where $V_n(\psi)$ are the Laurent modes of the vertex operator
\begin{equation} \label{eq:Niemeier:untwistedLaurent}
    V(\psi, z) = \sum_{n \in \mb Z} V_n (\psi) z^{-n - h_\psi}
\end{equation}
and $h_\psi$ the conformal dimension of $\psi$.

\subsection{The twisted sector}
For any lattice $\Lambda$, the theory we have described admits a canonical $\mb Z_2$ automorphism $\theta$, defined by
\begin{equation}
    \theta X^i(z) \theta^{-1} = - X^i(z),
\end{equation}
or
\begin{equation} \label{eq:Niemeier:untwistedInvolution}
    \theta a^i_n \theta^{-1} = -a^i_n \qq{and} \theta \ket{\lambda} = \ket{-\lambda}.
\end{equation}
Orbifolding by this canonical automorphism and taking the spin lift (in the manner described in the introduction to this section) of the resultant theory yields a canonically enlarged theory which we call the $\mb Z_2$ twisted VOA. The twisted sector of the theory contains $2^{r/2}$ ground states of conformal weight $\frac{r}{16}$, which furnish an irreducible representation of the Clifford algebra $\Gamma(\Lambda) = \qty{\pm \gamma_\lambda | \lambda \in \Lambda/2 \Lambda}$, with
\begin{equation} \label{eq:Niemeier:gammaCocycle}
    \gamma_\lambda \gamma_\mu = (-1)^{\lambda \cdot \mu} \gamma_\mu \gamma_\lambda \equiv \epsilon(\lambda, \mu) \gamma_{\lambda + \mu}.
\end{equation}
We call the carrying space of this representation $\mc{S}(\Lambda)$. These ground states are acted upon by half-integer moded oscillators $c_s$, $s \in \mb Z + \frac{1}{2}$, with 
\begin{equation} \label{eq:Niemeier:twistedOscillators}
    \comm{c_s^i}{c_t^j} = s \delta^{i j} \delta_{s, -t}, \qq{} c^{j \dag}_s = c^j_{-s}, \qq{and} c^j_s \chi = 0\ \forall\ s > 0,\ \chi \in \mc S(\Lambda).
\end{equation}
The operators
\begin{equation} \label{eq:Niemeier:twistedVirasoro}
    L^T_n = \frac{1}{2} \sum_{s \in \mb Z + \frac{1}{2}} \normord{c_s \cdot c_{n - s}} + \frac{r}{16} \delta_{n, 0}
\end{equation}
satisfy the Virasoro algebra (\ref{eq:Niemeier:VirasoroAlg}). The twisted sector Hilbert space $\mc H_\Lambda^T$ is spanned by the Fock spaces
\begin{equation}
    \chi = \qty(\prod_{a = 1}^M c_{-s_a}^{i_a}) \chi_0,
\end{equation}
where $s_a \in \mb Z^{\geq 0} + \frac{1}{2}$, distinct $s_a$ may coincide, and $\chi_0 \in \mc S(\Lambda)$. The state $\chi$ has conformal dimension
\begin{equation} \label{eq:Niemeier:twistedConformalDimension}
    L^T_0 \chi = \qty(\frac{r}{16} + \sum_{a = 1}^M s_a) \chi.
\end{equation}
There are also twisted sector fields
\begin{equation} \label{eq:Niemeier:twistedWorldsheetFields}
    R^i(z) = i \sum_{s \in \mb Z + \frac{1}{2}} \frac{c^i_s}{s} z^{-s}
\end{equation}
playing the role of the worldsheet fields $X^i(z)$ in the untwisted sector.

With these definitions, we may write down a set of vertex operators over the twisted sector which furnish a representation of the untwisted VOA. The twisted sector vertex operator associated to the state $\psi$ of (\ref{eq:Niemeier:genericState}) is
\begin{equation} \label{eq:Niemeier:twistedVertexOp}
    V_T(\psi, z) = V^0_T(e^{\Delta(z)} \psi, z)
\end{equation}
where
\begin{equation} \label{eq:Niemeier:twistedVertexOpGuess}
    V^0_T(\psi, z) = (4 z)^{-\lambda^2/2} \normord{\qty(\prod_{a = 1}^M \frac{i}{(n_a - 1)!} \dv[n_a]{R^{i_a}(z)}{z}) \exp(i \lambda \cdot R(z))} \gamma_\lambda
\end{equation}
and
\begin{equation}
    \Delta(z) = \frac{1}{2} \sum_{\substack{m, n \geq 0 \\ (m, n) \neq (0, 0)}} \binom{-\frac{1}{2}}{m} \binom{-\frac{1}{2}}{n} \frac{z^{-m - n}}{m + n} a_m \cdot a_n.
\end{equation}
As in the untwisted sector, the stress tensor is the vertex operator associated to the conformal state:
\begin{eqnarray}
    \sum_{n \in \mb Z} L^T_n z^{-n - 2} = V_T(\psi_L, z).
\end{eqnarray}\
Once again, we refer the reader to \cite{dolan_conformal_1990} for the details of the derivation, including a proof of mutual locality for the twisted sector vertex operators.

So far, we have seen that the twisted sector furnishes a representation of the untwisted VOA. To form the $\mb Z_2$ twisted theory, we wish to extend the untwisted VOA by this representation. This amounts to defining intertwiners, or vertex operators associated to the twisted sector states which map $\mc H_\Lambda$ to $\mc H^T_\Lambda$ and vice versa. We do so by writing
\begin{equation} \label{eq:Niemeier:interwtwinerDefinition}
    W(\chi, z) \psi = e^{z L^T_{-1}} V_T(\psi, -z) \chi
\end{equation}
and
\begin{equation}
    \overline{W}(\chi, z) = z^{-2 h_\chi} W \qty(e^{z^* L_1^T} \overline{\chi}, \frac{1}{z^*})^\dag,
\end{equation}
where $\overline{\chi}$ is the conjugate state to $\chi$ as defined in equation 2.4 of \cite{dolan_conformal_1990}. Then, writing a generic state in the $\mb Z_2$ twisted theory as a two-component vector in $\mc H_\Lambda \oplus \mc H^T_\Lambda$, the vertex operators associated to untwisted sector states and twisted sector states respectively are
\begin{equation}
    \tilde{V}(\psi, z) = \mqty(V(\psi, z) & 0 \\ 0 & V_T(\psi, z)) \qq{and} \tilde{V}(\chi, z) = \mqty(0 & \overline{W}(\chi, z) \\ W(\chi, z) & 0).
\end{equation}
In this language, the stress tensor for the full theory is diagonal:
\begin{equation}
    \tilde{T}(z) = \tilde{V}(\psi_L, z) = \mqty(V(\psi_L, z) & 0 \\ 0 & V_T(\psi_L, z)) = \sum_{n \in \mb Z} \mqty(L_n & 0 \\ 0 & L^T_n) z^{-n - 2} \equiv \tilde{L}_n z^{-n - 2}.
\end{equation}
A generic untwisted vertex operator $V(\psi, z)$ may be Laurent expanded as in (\ref{eq:Niemeier:untwistedLaurent}), but there are subtleties in the Laurent expansions of $V_T$, $W$, and $\overline{W}$. For the twisted vertex operators, we have
\begin{equation}
    V_T(\psi, z) = \sum_n (V_T(\psi))_n z^{-n - h_\psi}
\end{equation}
where the sum runs over $\mb Z$ (resp. $\mb Z + \frac{1}{2}$) if $\psi$ is in the $\theta = 1$ (resp. $\theta = -1$) subspace of the untwisted sector. For the intertwiners, the story is different depending on whether $\frac{r}{8}$ is odd or even; moving forward we take it to be odd (since $r = 24$ is of interest to this work). Then we may write
\begin{equation} \label{eq:Niemeier:intertwinerLaurent}
    \tilde{V}(\chi, z) = \sum_n \tilde{V}_n(\chi) z^{-n - h_\chi} = \sum_n \mqty(0 & \overline{W}_n(\chi) \\ \overline{W}_n(\chi) & 0) z^{-n - h_\chi}
\end{equation}
where the sum runs over $\mb Z$ (resp. $\mb Z + \frac{1}{2}$) if $\chi$ is in the Ramond (resp. Neveu-Schwarz) sector. It turns out (see below) that the VOA we have constructed is only superconformal and mutually local if we restrict to the NS sector, which amounts to projecting onto $\theta = 1$ in the untwisted sector and $\theta = -1$ in the twisted. Thus, moving forward we may take all twisted sector vertex operators to be integrally moded and all intertwiners half-integrally. With this understanding, equations (\ref{eq:Niemeier:L0WithV}) and (\ref{eq:Niemeier:LmWithVn}) may be applied to the twisted vertex operators and intertwiners.

The various types of vertex operators also satisfy so-called ``duality relations''
\begin{equation}
    \tilde{V}(\xi_1, z) \tilde{V}(\xi_2, w) = \tilde{V}(\tilde{V}(\xi_1, z - w) \xi_2, w)
\end{equation}
without restrictions on whether $\xi_{1, 2}$ are twisted sector states. Laurent expanding on both sides yields various relations between the modes of the (un)twisted vertex operators and the intertwiners; in particular, we will need that
\begin{equation} \label{eq:Niemeier:intertwinersDuality}
    \tilde{V}(\chi, z) \tilde{V}(\rho, w) = \sum_{n \in \mb Z + \frac{1}{2} + h_\rho} (z - w)^{n - h_\chi - h_\rho} \tilde{V} \qty(\overline{W}_{h_\rho - n}(\chi) \rho, w)
\end{equation}
and
\begin{equation} \label{eq:Niemeier:twistedIntertwinerDuality}
    \tilde{V}(\psi, z) \tilde{V}(\chi, w) = \sum_{n \in \mb Z + h_\chi} (z - w)^{n - h_\psi - h_\chi} \tilde{V} \qty((V_T(\psi))_{h_\chi - n} \chi, w),
\end{equation}
where $\psi$ is an untwisted sector state and $\chi$, $\rho$ are twisted.

The final step in constructing the theory of interest is to restrict to particular eigenspaces of the involution $\theta$. Specifically, the action of $\theta$ on the untwisted sector is as in (\ref{eq:Niemeier:untwistedInvolution}), and its action on the twisted sector is
\begin{equation} \label{eq:Niemeier:twistedInvolution}
    \theta c^i_s \theta^{-1} = -c^i_s \qq{and} \theta \chi_0 = (-1)^{r/8} \chi_0
\end{equation}
($\chi_0 \in \mc S(\Lambda)$). We refer to the $\theta = \pm 1$ subspaces of $\mc H_\Lambda$ and $\mc H^T_\Lambda$ as $\mc H^\pm_\Lambda \subset \mc H_\Lambda$ and $\mc H^{\pm, T}_\Lambda \subset \mc H^T_\Lambda$ respectively; then, the Hilbert space of the theory in question is
\begin{equation} \label{eq:Niemeier:SCVOAHilbertSpace}
    \tilde{\mc H}_\Lambda = \mc H^+_\Lambda \oplus \mc H^{-, T}_\Lambda.
\end{equation}
We pause to note that here we are following \cite{moore_beauty_2023}, rather than \cite{dolan_conformal_1990}, which chose $\mc H^+_\Lambda \oplus \mc H^{+, T}_\Lambda$ instead of (\ref{eq:Niemeier:SCVOAHilbertSpace}). The reason for this is twofold: firstly, the ``spin lift'' procedure described in the introduction to this section leads to the Hilbert space $\tilde{\mc H}_\Lambda$ \cite{kapustin_fermionic_2017, moore_beauty_2023}; moreover, the theory we have constructed is only an SVOA if we restrict to the NS sector, since including the R sector results in OPEs which violate the mutual locality condition (\ref{eq:Niemeier:mutualLoc}). One may show that the NS sector of this theory is precisely (\ref{eq:Niemeier:SCVOAHilbertSpace}) (cf. \cite[eq. 5.31]{moore_beauty_2023}). Secondly, we are interested in theories with $r = 24$, so using (\ref{eq:Niemeier:twistedInvolution}), the twisted sector ground states in our theory have $\theta = (-1)^{24/8} = -1$. As we will see in the next section, these states are necessary for superconformal symmetry, so our Hilbert space must include $\mc H^{-, T}_\Lambda$. When $\Lambda$ is chosen to be the Leech lattice, the theory resulting from this procedure is an SCVOA within the ``Beauty and the Beast'' theory \cite{dixon_beauty_1988, moore_beauty_2023}.

\section{Constructing supercurrents}
In this section, we show that the problem of constructing a supercurrent in a spin lift of a $\mb Z_2$ orbifold of a Niemeier theory reduces to the problem of finding a sublattice of a certain type in the associated Niemeier lattice. We will closely follow the derivation in section 5.2 of \cite{moore_beauty_2023}, deviating only where specific properties of the Leech lattice were used; those portions of the calculation which are identical to the Leech lattice case we will lay out only schematically and refer the reader to \cite{moore_beauty_2023} for details.

The idea of the construction is as follows: we notice that for the Niemeier lattices, $r = 24$, so the conformal weight of the twisted sector ground states spanning $\mc S(\Lambda)$ is $\frac{24}{16} = \frac{3}{2}$, the correct dimension for a supercurrent. We will consider the vertex operator associated to an arbitrary twisted sector ground state and compute its OPEs with the stress tensor and with itself, then show that by a careful choice of twisted sector ground state we can ensure that the OPEs are those of a supercurrent.

To begin, choose an arbitrary twisted sector ground state $\chi \in \mc S(\Lambda)$. We know from (\ref{eq:Niemeier:twistedConformalDimension}) that $h_\chi = \frac{3}{2}$. From (\ref{eq:Niemeier:twistedIntertwinerDuality}), the OPE of the associated vertex operator with the stress tensor is
\begin{equation}
    \begin{aligned}
        \tilde{V}(\psi_L, z) \tilde{V}(\chi, w) ={}& \sum_{n \in \mb Z + h_\chi} (z - w)^{n - h_{\psi_L} - h_\chi} \tilde{V} \qty((V_T(\psi_L))_{h_\chi - n} \chi, w) \\
        ={}& \sum_{n \in \mb Z + \frac{1}{2}} (z - w)^{n - 2 - \frac{3}{2}} \tilde{V} \qty(L^T_{\frac{3}{2} - n} \chi, w) \\
        ={}& \frac{\tilde{V} \qty(L_0^T \chi, w)}{(z - w)^2} + \frac{\tilde{V} \qty(L^T_{-1} \chi, w)}{z - w} + \text{reg.} \\
        ={}& \frac{3/2}{(z - w)^2} \tilde{V}(\chi, w) + \frac{1}{z - w} \dv{\tilde{V}(\chi, w)}{w} + \text{reg.}
    \end{aligned}
\end{equation}
where we have used (\ref{eq:Niemeier:twistedOscillators}) and (\ref{eq:Niemeier:twistedVirasoro}) to see that $\chi$ is annihilated by $L^T_n$ for all $n > 0$. The fact that $\tilde{V}(L^T_{-1} \chi, w) = \dv{\tilde{V}(\chi, w)}{w}$ is a consequence of (\ref{eq:Niemeier:definingPropBrokenOut}) and the uniqueness property discussed below it. Thus, $\tilde{V}(\chi, w)$ is a conformal primary for any $\chi \in \mc S(\Lambda)$. If we can choose $\chi$ such that
\begin{equation} \label{eq:supercurrent:supercurrentOPE}
    \tilde{V}(\chi, z) \tilde{V}(\chi, w) = \frac{4}{(z - w)^3} + \frac{1/2}{z - w} \tilde{V}(\psi_L, z) + \text{reg.}
\end{equation}
$V(\chi, z)$ will be a superconformal current. To this end, choose (potentially distinct) twisted sector ground states $\chi_a, \chi_b \in \mc S(\Lambda)$. We have
\begin{equation} \label{eq:supercurrent:distinctGSOPE}
    \begin{aligned}
        \tilde{V}(\chi_a, z) \tilde{V}(\chi_b, w) ={}& \tilde{V}(\tilde{V}(\chi_a, z - w) \chi_b, w) \\
        ={}& \tilde{V}(\overline{W}(\chi_a, z - w) \chi_b, w) \\
        ={}& \sum_{n \in \mb Z} (z - w)^{n - 3} \tilde{V} \qty(\overline{W}_{\frac{3}{2} - n}(\chi_a) \chi_b, w).
    \end{aligned}
\end{equation}
Using (\ref{eq:Niemeier:LmWithVn}),
\begin{equation} \label{eq:supercurrent:WchiConformalWeight}
    \begin{aligned}
        \comm{\tilde{L}_0}{\tilde{V}_n(\chi_a)} \chi_b ={}& -n \tilde{V}_n(\chi_a) \chi_b \\
        L_0 \overline{W}_n(\chi_a) \chi_b - \overline{W}_n(\chi_a) L^T_0 \chi_b ={}& -n \overline{W}_n(\chi_a) \chi_b \\
        L_0 \qty(\overline{W}_n(\chi_a) \chi_b) ={}& \qty(\frac{3}{2} - n) \qty(\overline{W}_n(\chi_a) \chi_b),
    \end{aligned}
\end{equation}
so the state $\overline{W}_n(\chi_a) \chi_b$ has conformal dimension $\frac{3}{2} - n$; then, since conformal weights are nonnegative, we must have
\begin{equation}
    \overline{W}_n(\chi_a) \chi_b = 0 \qq{} \forall\ n > \frac{3}{2}.
\end{equation}
This means that the sum over $\mb Z$ in (\ref{eq:supercurrent:distinctGSOPE}) may be restricted to $\mb Z^{\geq 0}$, so that
\begin{equation} \label{eq:supercurrent:VchiVchiOPEAnsatz}
    \tilde{V}(\chi, z) \tilde{V}(\chi, w) = \frac{\tilde{V} \qty(\overline{W}_{\frac{3}{2}}(\chi) \chi, w)}{(z - w)^3} + \frac{\tilde{V} \qty(\overline{W}_{\frac{1}{2}}(\chi) \chi, w)}{(z - w)^2} + \frac{\tilde{V} \qty(\overline{W}_{-\frac{1}{2}}(\chi) \chi, w)}{z - w} + \text{reg.}
\end{equation}
(where we have also set $\chi_a = \chi_b = \chi$). We see that the relevant states are $\overline{W}_s(\chi) \chi$ with $s \in \qty{\frac{3}{2}, \frac{1}{2}, -\frac{1}{2}}$. From (\ref{eq:supercurrent:WchiConformalWeight}) these have conformal weights $\qty{0, 1, 2}$ respectively. Thus, using (\ref{eq:Niemeier:untwistedConformalDimension}), we have the ans\"atze
\begin{align}
    \overline{W}_{\frac{3}{2}}(\chi) \chi ={}& \alpha(\chi) \ket{0}, \label{eq:supercurrent:W3/2Ansatz}\\
    \overline{W}_{\frac{1}{2}}(\chi) \chi ={}& \sum_i \nu_i a^i_{-1} \ket{0} + \sum_{\lambda^2 = 2} \nu(\lambda) \ket{\lambda}, \label{eq:supercurrent:W1/2Ansatz} \\
    \overline{W}_{-\frac{1}{2}}(\chi) \chi ={}& \sum_i \delta_i a^i_{-1} a^i_{-1} \ket{0} + \sum_{i < j} \kappa_{i j} a^i_{-1} a^j_{-1} \ket{0} + \sum_j \kappa_j a^j_{-2} \ket{0} \label{eq:supercurrent:W-1/2Ansatz} \\
    {}& \hspace{3cm} + \sum_{\lambda^2 = 4} \kappa(\lambda) \ket{\lambda} + \sum_{j, \lambda^2 = 2} \mu_j(\lambda) a^j_{-1} \ket{\lambda}. \nonumber
\end{align}
Here we have departed from \cite{moore_beauty_2023}; since that work focused on the Leech lattice, their ans\"atze (equations 5.56 and 5.60) did not include the terms with $\lambda^2 = 2$ in (\ref{eq:supercurrent:W1/2Ansatz}) and (\ref{eq:supercurrent:W-1/2Ansatz}). Taking the inner product between $\ket{0}$ and (\ref{eq:supercurrent:W3/2Ansatz}) gives
\begin{equation}
    \begin{aligned}
        \alpha(\chi) ={}& \mel{0}{\overline{W}_{\frac{3}{2}}(\chi)}{\chi} \\
        ={}& \qty(W_{-\frac{3}{2}}(\overline{\chi}) \ket{0})^\dag \ket{\chi}
    \end{aligned}
\end{equation}
but by (\ref{eq:Niemeier:interwtwinerDefinition}) and (\ref{eq:Niemeier:intertwinerLaurent}) the term in brackets is just the constant part of $e^{z L^T_{-1}} \overline{\chi}$, so
\begin{equation} \label{eq:supercurrent:alphachi}
    \alpha(\chi) = \ip{\overline{\chi}}{\chi}.
\end{equation}
Next, using (\ref{eq:supercurrent:alphachi}), (\ref{eq:Niemeier:interwtwinerDefinition}), and (\ref{eq:Niemeier:LmWithVn}) with $(m, n) = \qty(2, -\frac{1}{2})$, we have
\begin{equation}
    \mel{0}{L_2 \overline{W}_{-\frac{1}{2}}(\chi)}{\chi} = \frac{3}{2} \alpha(\chi).
\end{equation}
Using (\ref{eq:supercurrent:W-1/2Ansatz}), this becomes
\begin{equation}
    \begin{aligned}
        \frac{3}{2} \alpha(\chi) ={}& \sum_i \delta_i \mel{0}{L_2 a^i_{-1} a^i_{-1}}{0} + \sum_{i < j} \kappa_{i j} \mel{0}{L_2 a^i_{-1} a^j_{-1}}{0} + \sum_j \kappa_j \mel{0}{L_2 a^j_{-2}}{0} \\
    {}& \hspace{4cm} + \sum_{\lambda^2 = 4} \kappa(\lambda) \mel{0}{L_2}{\lambda} + \sum_{j, \lambda^2 = 2} \mu_j(\lambda) \mel{0}{L_2 a^j_{-1}}{\lambda}
    \end{aligned}
\end{equation}
but from (\ref{eq:Niemeier:untwistedStressTensor}) and the oscillator algebra (\ref{eq:Niemeier:untwistedOscillators}) all the matrix elements on the RHS vanish except
\begin{equation}
    \mel{0}{L_2 a^i_{-1} a^i_{-1}}{0} = \delta^{i j},
\end{equation}
so
\begin{equation}
    \frac{3}{2} \alpha(\chi) = \sum_i \delta_i.
\end{equation}
The term in (\ref{eq:supercurrent:W-1/2Ansatz}) containing $\kappa_j$ has the wrong parity under the reflection $\theta$ and must vanish (cf. equations 5.87 - 5.92 in \cite{moore_beauty_2023}), and to determine the remaining coefficients in (\ref{eq:supercurrent:W-1/2Ansatz}), we will use the facts that
\begin{gather}
    \mel{0}{a^m_1 a^n_1 \overline{W}_{-\frac{1}{2}}(\chi)}{\chi} = \begin{cases}
        \kappa_{n m} & n \neq m \\
        2 \delta_m & n = m
    \end{cases}, \\
    \mel{\lambda}{\overline{W}_{-\frac{1}{2}}(\chi)}{\chi} = \kappa(\lambda), \\
    \mel{\lambda}{a^j_1 \overline{W}_{-\frac{1}{2}}(\chi)}{\chi} = \mu_j(\lambda),
\end{gather}
all of which are consequences of the ansatz (\ref{eq:supercurrent:W-1/2Ansatz}) and the oscillator algebra (\ref{eq:Niemeier:untwistedOscillators}). The general approach going forward is to use the fact that
\begin{equation} \label{eq:supercurrent:generalApproach}
    \begin{aligned}
        \mel{\psi}{\overline{W}_{-\frac{1}{2}}(\chi)}{\chi} ={}& \qty(\mel{\chi}{W_{\frac{1}{2}}(\overline{\chi})}{\psi})^* \\
        ={}& \text{coefficient of } \bar{z}^{-2} \text{ in } \qty(\mel{\chi}{e^{z L^T_{-1}} V_T(\psi, -z)}{\overline{\chi}})^*
    \end{aligned}
\end{equation}
and evaluate $V_T(\psi, -z)$ by means of (\ref{eq:Niemeier:twistedVertexOp}). For $\delta_m$, $\kappa_{n m}$, and $\kappa(\lambda)$, the calculation proceeds exactly as in \cite{moore_beauty_2023}, and we obtain
\begin{equation} \label{eq:supercurrent:coefficientResults}
    \kappa_{n m} = 0, \qq{} \delta_i = \frac{\alpha(\chi)}{16}, \qq{} \kappa(\lambda) = 16 \mel{\chi}{e_\lambda}{\overline{\chi}}^*.
\end{equation}
To compute $\mu_j(\lambda)$, we apply (\ref{eq:supercurrent:generalApproach}) with $\psi = a^j_{-1} \ket{\lambda}$, $\lambda^2 = 2$. From (\ref{eq:Niemeier:twistedVertexOp}),
\begin{equation} \label{eq:supercurrent:VTofalambda}
    V_T \qty(a^j_{-1} \ket{\lambda}, -z) = V^0_T \qty(e^\Delta(z) a^j_{-1} \ket{\lambda}, -z).
\end{equation}
We have
\begin{equation} \label{eq:supercurrent:eDeltaonalambda}
    e^{\Delta(z)} a^j_{-1} \ket{\lambda} = \exp(\frac{1}{2} \sum_{\substack{m, n \geq 0 \\ (m, n) \neq (0, 0)}} A^{m n} a_m \cdot a_n - \frac{1}{2} a_0 \cdot a_0 \ln(4 z)) a^j_{-1} \ket{\lambda}
\end{equation}
with
\begin{equation}
    A^{m n} = \binom{-\frac{1}{2}}{m} \binom{-\frac{1}{2}}{n} \frac{(-z)^{-m - n}}{m + n}.
\end{equation}
Note that $a_0^i = p^i$ commutes with everything but $q^i$, which does not appear in (\ref{eq:supercurrent:eDeltaonalambda}), so $a_0 \cdot a_0$ acts on $\ket{\lambda}$ and becomes $\lambda^2$. Next, consider a generic term in the Taylor expansion of (\ref{eq:supercurrent:eDeltaonalambda}). If the term contains any annihilation operators $a_n$ with $n \geq 2$, consider the rightmost such $a_n$. It commutes freely with everything to its right and so annihilates against $\ket{\lambda}$, leaving
\begin{equation}
    \begin{aligned}
        e^{\Delta(z)} a^j_{-1} \ket{\lambda} ={}& \exp(\frac{1}{2} \qty(A^{1 0} a_1 \cdot a_0 + A^{0 1} a_0 \cdot a_1 + A^{11} a_1 \cdot a_1 - \lambda^2 \ln(4 z))) a^j_{-1} \ket{\lambda} \\
        ={}& \exp(\frac{1}{2} \qty(\qty(A^{01} + A^{10}) \lambda \cdot a_1 - \lambda^2 \ln(4 z))) a^j_{-1} \ket{\lambda} \\
        ={}& (4 z)^{-\frac{\lambda^2}{2}} \qty(1 + \frac{1}{2} \qty(A^{1 0} + A^{0 1}) \lambda \cdot a_1) a^j_{-1} \ket{\lambda}
    \end{aligned}
\end{equation}
where we have used that $a_0 \ket{\lambda} = \lambda$ and $a^j_1 a^k_1$ annihilates $a^i_{-1} \ket{\lambda}$ for all $j$, $k$. But
\begin{equation}
    a^i_1 a^j_{-1} \ket{\lambda} = \delta^{i j} \ket{\lambda},
\end{equation}
so (\ref{eq:supercurrent:VTofalambda}) becomes
\begin{equation}
    V_T \qty(a^j_{-1} \ket{\lambda}, -z) = (4 z)^{-\frac{\lambda^2}{2}} V^0_T \qty(a^j_{-1} \ket{\lambda}, -z) + (4 z)^{-\frac{\lambda^2}{2}} \qty(\frac{A^{1 0} + A^{0 1}}{2}) \lambda^j V^0_T \qty(\ket{\lambda}, -z).
\end{equation}
Using (\ref{eq:Niemeier:twistedWorldsheetFields}) and (\ref{eq:Niemeier:twistedVertexOpGuess}), the first term is
\begin{equation}
    \begin{aligned}
        (4 z)^{-\frac{\lambda^2}{2}} V^0_T(a^j_{-1} \ket{\lambda}, -z) ={}& (4 z)^{-\frac{\lambda^2}{2}} i \normord{\dv{R^j}{z}(-z) \exp(i \lambda \cdot R(-z))} \gamma_\lambda \\
        ={}& \frac{1}{4 z} \sum_{k \in \mb Z + \frac{1}{2}} \normord{c_k^j (-z)^{-k - 1} \exp(i \lambda \cdot R(-z))} \gamma_\lambda
    \end{aligned}
\end{equation}
and the second is
\begin{equation}
    (4 z)^{-\frac{\lambda^2}{2}} \qty(\frac{A^{1 0} + A^{0 1}}{2}) \lambda^j V^0_T \qty(\ket{\lambda}, -z) = \frac{1}{4 z} \qty(\frac{A^{1 0} + A^{0 1}}{2}) \lambda^j \normord{\exp(i \lambda \cdot R(-z))} \gamma_\lambda
\end{equation}
where we have made use of the fact that $\lambda^2 = 2$. Using (\ref{eq:supercurrent:generalApproach}), $\mu^*_j(\lambda)$ is the coefficient of $z^{-2}$ in
\begin{equation} \label{eq:supercurrent:powerCounting}
    \begin{aligned}
        {}& \mel{\chi}{e^{z L^T_{-1}} V_T(a^j_{-1} \ket{\lambda}, -z)}{\overline{\chi}} \\
        {}& \hspace{0.5cm} = \mel{\chi}{e^{z L^T_{-1}} \qty(\frac{1}{4 z}) \normord{\qty(\qty(\frac{A^{1 0} + A^{0 1}}{2}) \lambda^j + \sum_{k \in \mb Z + \frac{1}{2}} c^j_k (-z)^{-k - 1}) \exp(i \lambda \cdot R(-z))} \gamma_\lambda}{\overline{\chi}}.
    \end{aligned}
\end{equation}
The normal ordering means that any term containing annihilation operators will annihilate against $\gamma_\lambda \ket{\overline{\chi}}$; thus, the sum over $k$ may be restricted to $k < 0$, and similarly we may replace $R^j(-z)$ by
\begin{equation}
    R^j_-(-z) = i \sum_{\substack{r \in \mb Z + \frac{1}{2} \\ r < 0}} \frac{c^j_r}{r} z^{-r}.
\end{equation}
But then the most negative possible power of $z$ appearing on the RHS of (\ref{eq:supercurrent:powerCounting}) is when $k = r = 0$, which gives $(-z)^{-1}$. Therefore, $z^{-2}$ does not appear in (\ref{eq:supercurrent:powerCounting}), so $\mu_j(\lambda) = 0$. Taken together with (\ref{eq:supercurrent:coefficientResults}), and using (\ref{eq:Niemeier:conformalStateExplicit}), this means that (\ref{eq:supercurrent:W-1/2Ansatz}) reduces to
\begin{equation}
    \overline{W}_{-\frac{1}{2}}(\chi) \chi = \frac{\alpha(\chi)}{8} \psi_L + \sum_{\lambda^2 = 4} \kappa(\lambda) \ket{\lambda}.
\end{equation}
Applying (\ref{eq:Niemeier:LmWithVn}) with $(m, n) = \qty(1, -\frac{1}{2})$ and acting on $\chi$ with the result yields
\begin{equation}
    \overline{W}(\chi)_{\frac{1}{2}} (\chi) \chi = L_1 \overline{W}_{-\frac{1}{2}}(\chi) \chi = \frac{\alpha(\chi)}{8} L_1 \psi_L + \sum_{\lambda^2 = 4} \kappa(\lambda) L_1 \ket{\lambda} = 0.
\end{equation}
All told, then, we have simplified (\ref{eq:supercurrent:VchiVchiOPEAnsatz}) to
\begin{equation}
    \tilde{V}(\chi, z) \tilde{V}(\chi, w) = \frac{\alpha(\chi)}{(z - w)^3} + \frac{\alpha(\chi)/8}{z - w} T(z) + \sum_{\lambda^2 = 4} \frac{\kappa(\lambda)}{z - w}
\end{equation}
which is precisely (5.93) of \cite{moore_beauty_2023}. To obtain a superconformal current, we need only find $\chi$ such that $\kappa(\lambda)$ vanishes for all $\lambda^2 = 4$. This portion of the derivation proceeds exactly as in \cite{moore_beauty_2023}, and so we review it schematically. Choose an order $2^{12}$ Abelian subgroup $\Gamma^{12} < \Gamma(\Lambda)$ such that $-1 \notin \Gamma^{12}$, then choose a basis for $\Gamma(\Lambda)$ which trivializes the cocycle in (\ref{eq:Niemeier:gammaCocycle}) over $\Gamma^{12}$. We may then form the rank-one projection operator
\begin{equation} \label{eq:supercurrent:projector}
    P_{12}(\Gamma^{12}) = \frac{1}{2^{12}} \sum_{\gamma \in \Gamma^{12}} \gamma.
\end{equation}
Define a ``superconformal sublattice'' to be a sublattice $\Lambda_{SC} < \Lambda$ with the following properties:
\begin{enumerate}
    \item $\Lambda_{SC}/\sqrt{2}$ is even and integral.
    \item $\Lambda_{SC}$ has minimum squared-norm greater than 4.
    \item $2 \Lambda < \Lambda_{SC}$ with index $2^{12}$.
\end{enumerate}
Then the group $\Gamma_{SC}(\Lambda_{SC}) = \qty{\gamma_\lambda\ |\ \lambda \in \Lambda_{SC}}$ is an order $2^{12}$ Abelian subgroup of $\Gamma(\Lambda)$ not containing $-1$, and so we may construct the projector $P_{12}$ of (\ref{eq:supercurrent:projector}). The image of this operator contains a unique vector $\chi_1$ up to normalization, and this vector has the property that $\kappa(\lambda) = 0$ for all $\lambda \in \Lambda$ with $\lambda^2 = 4$ \cite[Theorem 5.2]{moore_beauty_2023}. Thus, given a superconformal sublattice $\Lambda_{SC} < \Lambda$, the operator
\begin{equation}
    2 \tilde{V}(\chi_1, z), \qq{} \chi_1 \in \Im P_{12}(\Gamma_{SC}(\Lambda_{SC}))
\end{equation}
satisfies the OPE (\ref{eq:supercurrent:supercurrentOPE}) and is a superconformal current in the $\mb Z_2$ twisted Niemeier theory associated to $\Lambda$.

\section{Superconformal sublattices}
We have seen that superconformal sublattices of the Niemeier lattices give superconformal currents in the associated CFTs. It remains to check which of the Niemeier lattices admit such sublattices.\footnote{We thank Frank Calegari for some extremely helpful advice on efficiently finding superconformal sublattices.}
As a first approach, notice that all superconformal sublattices are isomorphic to $\sqrt{2}$ times the Leech lattice, since by definition $\Lambda_{SC}/\sqrt{2}$ must be even, unimodular, and have minimum squared-norm greater than 2. This suggests a procedure for finding superconformal sublattices: look for isometric embeddings of the form
\begin{equation}\label{eq:SCSLs:embedding}
    \Lambda_L > \sqrt{2} \Lambda > 2 \Lambda_L
\end{equation}
with $\Lambda_L$ the Leech lattice and $\Lambda$ another Niemeier lattice. Dividing through by $\sqrt{2}$, this induces an embedding $\Lambda > \sqrt{2} \Lambda_L$, the image of which is then a superconformal sublattice of $\Lambda$. Embeddings of this type have been considered before, for example in \cite{dong_associative_1996}, where it was shown that such embeddings exist for all Niemeier lattices. In principle, this suffices to demonstrate that all spin lifted $\mb Z_2$ orbifolds of Niemeier theories enjoy supersymmetry, but without finding the SCSLs, one cannot explicitly construct the supercurrents. Randomly seeking embeddings of the form (\ref{eq:SCSLs:embedding}) is very inefficient, since some Niemeier lattices permit many more such embeddings than others (cf. Table 1 of \cite{dong_associative_1996}), and in fact it would take prohibitive amounts of compute time to find such embeddings for all Niemeier lattices. Fortunately, we can take a different approach, and explicitly seek SCSLs within the Niemeier lattices themselves. Our algorithm for doing so is straightforward:
\begin{enumerate}
    \item Given a Niemeier lattice $\Lambda$, consider $\tilde{\Lambda} = \Lambda/2 \Lambda$ (thought of as an inner product space over $\mb F_2^{24}$). Choose a random vector $v$ whose squared-norm is 0 mod 4, but greater than 4, in $\tilde{\Lambda}$, and let $\tilde{\Lambda}_{SC}$ be the subspace $\qty{0, v}$.
    \item Compute $\tilde{\Lambda}_{SC}^\perp$, the orthogonal complement of $\tilde{\Lambda}_{SC}$ in $\tilde{\Lambda}$.
    \item Choose a random $u \in \tilde{\Lambda}_{SC}^\perp$ with squared-norm 0 mod 4, and add it to $\tilde{\Lambda}_{SC}$.
    \item Check that the minimum squared-norm in $\tilde{\Lambda}_{SC}$ is still greater than 4. If not, start over from step 1.
    \item Repeat steps 2-4 until $\tilde{\Lambda}_{SC}$ is a dimension 12 subspace of $\tilde{\Lambda}$.
\end{enumerate}
$\Lambda_{SC} = \tilde{\Lambda}_{SC} + 2 \Lambda$ is then a superconformal sublattice of $\Lambda$. While this procedure can fail at step 4, in practice it works sufficiently often that the algorithm terminates quickly in most cases. Implementing this approach in Magma, we find that all 24 of the Niemeier lattices admit superconformal sublattices. An example superconformal sublattice of the $A_2^{12}$ Niemeier lattice is given in appendix \ref{ss:exampleLattice}; examples for the remaining Neimeier lattices may be found in the supplementary data associated to this paper on the arXiv. This means that all $\mb Z_2$ twisted Niemeier theories enjoy $\mc N = 1$ SUSY; a fact which was known in some cases, such as for the Leech lattice \cite{dixon_beauty_1988, moore_beauty_2023}, the $A_1^{24}$ lattice \cite{benjamin_extremal_2015}, and the $A_2^{12}$ lattice \cite{harrison_extremal_2016}, but not systematically. Moreover, one may use these superconformal sublattices to explicitly construct the associated supercurrents.

\section{Conclusions}
We have seen that, in an extension of the approach taken by \cite{moore_beauty_2023} in the context of the Beauty and the Beast theory, one may construct $\mc N = 1$ supercurrents in $\mb Z_2$ twisted Niemeier theories using special sublattices of the associated Niemeier lattices. By a direct search, we have found superconformal sublattices of all 24 Niemeier lattices, meaning that all $\mb Z_2$ twisted Niemeier theories enjoy supersymmetry. Several directions of potentially promising future work present themselves:
\begin{itemize}
    \item In \cite{hohn_systematic_2022} it was shown that all 70 of the Schellekens theories at $c = 24$ may be constructed as $\mb Z_n$ orbifolds of the Niemeier theories. It is natural to ask whether our construction can be extended to the case $n > 2$, potentially using a different type of sublattice.
    \item Some of the $\mb Z_2$ twisted Niemeier theories are known to possess higher SUSY: in particular, the $A_1^{24}$ theory has $\mc N = 2$ and the $A_2^{12}$ theory $\mc N = 4$ \cite{benjamin_extremal_2015,harrison_extremal_2016}. The Beauty and the Beast theory cannot have $\mc N > 1$ SUSY since it contains no currents (equivalently, the Leech lattice contains no elements with squared-norm 2), but this is not true of the other Niemeier lattices. Since superconformal sublattices which are not related by the action of the lattice automorphism group give rise to distinct supercurrents under our construction, it seems plausible that some of the other $\mb Z_2$ twisted Niemeier theories may permit higher SUSY.
    \item The aforementioned $A_1^{24}$ and $A_2^{12}$ theories are so-called ``extremal,'' in that they have the largest gap in the spectrum of conformal weights permitted by the Virasoro algebra at $c = 24$ and (higher) SUSY. It would be interesting to seek an understanding of this extremality in terms of properties of the underlying lattices.
\end{itemize}

\section{Acknowledgements}
This work was supported by the National Science Foundation grant PHY-2310635.

\appendix
\section{An example superconformal sublattice} \label{ss:exampleLattice}
Here we present an explicit example of a superconformal sublattice in one of the Niemeier lattices. Adopting (non-orthonormal) coordinates for the $A_2^{12}$ Niemeier lattice such that the Gram matrix (obtained from Nebe and Sloane's catalogue of lattices \cite{nebe_catalogue_nodate}) is
\begin{equation}
    \smqty(2 & 0 & 0 & 0 & 0 & 0 & 0 & 0 & 0 & 1 & 1 & 1 & 1 & 1 & 1 & 1 & 1 & 0 & 0 & 0 & 0 & 0 & 0 & 0\\
    0 & 2 & 0 & 0 & 0 & 0 & 0 & 0 & 0 & 1 & 1 & 1 & 1 & -1 & 0 & 0 & 0 & 1 & 1 & 1 & 1 & 1 & 1 & 1\\
    0 & 0 & 2 & 0 & 0 & 0 & 0 & 0 & 0 & 1 & 0 & 0 & 0 & -1 & 0 & -1 & -1 & 1 & 1 & -1 & -1 & 0 & 1 & 0\\
0 & 0 & 0 & 2 & 0 & 0 & 0 & 0 & 0 & 0 & 1 & -1 & 1 & -1 & 1 & -1 & 0 & 1 & -1 & 1 & 0 & -1 & -1 & 0\\
0 & 0 & 0 & 0 & 2 & 0 & 0 & 0 & 0 & 0 & 1 & 0 & 0 & 0 & -1 & -1 & 1 & 0 & -1 & -1 & -1 & -1 & 0 & 1\\
0 & 0 & 0 & 0 & 0 & 2 & 1 & 0 & 0 & -1 & 0 & 0 & 1 & 1 & -1 & 1 & 1 & -1 & -1 & -1 & 1 & -1 & -1 & -1\\
0 & 0 & 0 & 0 & 0 & 1 & 2 & 0 & 0 & -1 & 0 & 0 & 1 & 1 & -1 & 0 & 0 & 0 & -1 & -1 & 1 & -1 & -1 & 0\\
0 & 0 & 0 & 0 & 0 & 0 & 0 & 2 & -1 & 1 & -1 & -1 & 1 & 0 & 1 & -1 & -1 & -1 & 0 & 0 & 1 & 1 & -1 & 1\\
0 & 0 & 0 & 0 & 0 & 0 & 0 & -1 & 2 & -1 & 0 & 1 & -1 & 0 & 0 & 1 & 1 & 1 & 0 & 0 & -1 & -1 & 0 & -1\\
1 & 1 & 1 & 0 & 0 & -1 & -1 & 1 & -1 & 4 & 1 & 1 & 1 & -1 & 2 & -1 & -1 & 1 & 2 & 1 & 0 & 2 & 1 & 1\\
1 & 1 & 0 & 1 & 1 & 0 & 0 & -1 & 0 & 1 & 4 & 1 & 1 & 0 & 0 & 0 & 2 & 1 & 0 & 1 & 0 & -1 & 1 & 0\\
1 & 1 & 0 & -1 & 0 & 0 & 0 & -1 & 1 & 1 & 1 & 4 & 0 & 0 & 0 & 2 & 1 & 1 & 1 & 0 & 0 & 1 & 1 & -1\\
1 & 1 & 0 & 1 & 0 & 1 & 1 & 1 & -1 & 1 & 1 & 0 & 4 & 0 & 1 & 0 & 0 & 0 & -1 & 0 & 2 & 0 & -1 & 1\\
1 & -1 & -1 & -1 & 0 & 1 & 1 & 0 & 0 & -1 & 0 & 0 & 0 & 4 & -1 & 2 & 2 & -2 & -1 & -1 & 1 & -1 & -1 & -1\\
1 & 0 & 0 & 1 & -1 & -1 & -1 & 1 & 0 & 2 & 0 & 0 & 1 & -1 & 4 & 0 & -1 & 1 & 1 & 2 & 0 & 1 & -1 & 0\\
1 & 0 & -1 & -1 & -1 & 1 & 0 & -1 & 1 & -1 & 0 & 2 & 0 & 2 & 0 & 4 & 2 & -1 & 0 & 0 & 1 & 0 & 0 & -2\\
1 & 0 & -1 & 0 & 1 & 1 & 0 & -1 & 1 & -1 & 2 & 1 & 0 & 2 & -1 & 2 & 4 & -1 & -1 & 0 & 0 & -2 & 0 & -1\\
0 & 1 & 1 & 1 & 0 & -1 & 0 & -1 & 1 & 1 & 1 & 1 & 0 & -2 & 1 & -1 & -1 & 4 & 1 & 1 & -1 & 0 & 1 & 1\\
0 & 1 & 1 & -1 & -1 & -1 & -1 & 0 & 0 & 2 & 0 & 1 & -1 & -1 & 1 & 0 & -1 & 1 & 4 & 1 & 0 & 2 & 2 & 0\\
0 & 1 & -1 & 1 & -1 & -1 & -1 & 0 & 0 & 1 & 1 & 0 & 0 & -1 & 2 & 0 & 0 & 1 & 1 & 4 & 1 & 1 & 0 & 0\\
0 & 1 & -1 & 0 & -1 & 1 & 1 & 1 & -1 & 0 & 0 & 0 & 2 & 1 & 0 & 1 & 0 & -1 & 0 & 1 & 4 & 1 & -1 & 0\\
0 & 1 & 0 & -1 & -1 & -1 & -1 & 1 & -1 & 2 & -1 & 1 & 0 & -1 & 1 & 0 & -2 & 0 & 2 & 1 & 1 & 4 & 1 & 1\\
0 & 1 & 1 & -1 & 0 & -1 & -1 & -1 & 0 & 1 & 1 & 1 & -1 & -1 & -1 & 0 & 0 & 1 & 2 & 0 & -1 & 1 & 4 & 1\\
0 & 1 & 0 & 0 & 1 & -1 & 0 & 1 & -1 & 1 & 0 & -1 & 1 & -1 & 0 & -2 & -1 & 1 & 0 & 0 & 0 & 1 & 1 & 4),
\end{equation}
the columns of the following matrix span a superconformal sublattice:
\begin{equation}
    \smqty(0 & 1 & 0 & 0 & 0 & 0 & 0 & 0 & 0 & 0 & -1 & 0 & 1 & 0 & 0 & 0 & 0 & 0 & 0 & 0 & 0 & 0 & 0 & 0 \\
2 & 0 & 0 & 0 & 0 & 0 & 0 & 0 & 0 & 0 & 0 & 0 & 0 & 0 & 0 & 0 & 0 & 0 & 0 & 0 & 0 & 0 & 0 & 0 \\
0 & 2 & 0 & 0 & 0 & 0 & 0 & 0 & 0 & 0 & 0 & 0 & 0 & 0 & 0 & 0 & 0 & 0 & 0 & 0 & 0 & 0 & 0 & 0 \\
0 & 1 & 0 & 0 & 0 & 0 & 0 & 0 & 0 & 0 & 0 & 0 & 0 & 0 & 0 & 0 & 1 & -1 & 0 & 0 & 0 & 0 & 0 & -1 \\
0 & 1 & 0 & 0 & 0 & 0 & 0 & 0 & 0 & 0 & 0 & 0 & 0 & 0 & 0 & 0 & -1 & 1 & 0 & 0 & 0 & 0 & 0 & -1 \\
0 & 1 & 0 & 0 & 0 & 0 & 0 & 0 & 0 & 0 & 0 & 0 & 0 & 0 & 0 & 0 & -1 & -1 & 0 & 0 & 0 & 0 & 0 & -1 \\
0 & 0 & 2 & 0 & 0 & 0 & 0 & 0 & 0 & 0 & 0 & 0 & 0 & 0 & 0 & 0 & 0 & 0 & 0 & 0 & 0 & 0 & 0 & 0 \\
0 & 0 & 0 & 2 & 0 & 0 & 0 & 0 & 0 & 0 & 0 & 0 & 0 & 0 & 0 & 0 & 0 & 0 & 0 & 0 & 0 & 0 & 0 & 0 \\
0 & 0 & 0 & 0 & 2 & 0 & 0 & 0 & 0 & 0 & 0 & 0 & 0 & 0 & 0 & 0 & 0 & 0 & 0 & 0 & 0 & 0 & 0 & 0 \\
0 & 0 & 0 & 0 & 0 & 0 & 2 & 0 & 0 & 0 & 0 & 0 & 0 & 0 & 0 & 0 & 0 & 0 & 0 & 0 & 0 & 0 & 0 & 0 \\
0 & 1 & 0 & 1 & 0 & 0 & 0 & 0 & 0 & 0 & 0 & 0 & 0 & -1 & -1 & 0 & 0 & 0 & 1 & 0 & 0 & 0 & 0 & 0 \\
0 & 0 & 1 & 0 & 0 & -1 & 0 & 1 & 0 & 0 & 0 & 0 & 0 & 0 & 0 & 0 & 0 & 0 & 1 & 0 & 0 & 0 & -1 & 0 \\
1 & 0 & 0 & 1 & 1 & 0 & -1 & 0 & 0 & 1 & 0 & 0 & 0 & 0 & 0 & 0 & 0 & 0 & 0 & 0 & 1 & 0 & 0 & 0 \\
0 & 0 & 1 & 1 & 0 & 0 & 0 & 0 & 0 & -1 & 0 & 0 & 0 & 0 & -1 & 0 & 0 & 0 & 0 & 0 & 0 & -1 & 0 & -1 \\
1 & 0 & -1 & 0 & 1 & 0 & 1 & 0 & 0 & 0 & 0 & 0 & 0 & 0 & -1 & 0 & 0 & 0 & 0 & 0 & 0 & 1 & -1 & 0 \\
1 & 0 & -1 & 0 & 1 & 0 & 0 & 0 & 0 & 0 & 1 & 0 & 0 & 0 & 1 & -1 & 1 & 0 & 0 & 0 & 0 & 0 & 0 & 0 \\
1 & 0 & 0 & 1 & 1 & 0 & 1 & 0 & 0 & 1 & 0 & 0 & 0 & 0 & 0 & 0 & 0 & 0 & 0 & 0 & 0 & 0 & -1 & -1 \\
1 & 0 & 0 & 1 & 1 & 0 & -1 & 0 & 0 & -1 & 0 & 0 & 0 & 0 & 0 & 0 & 0 & 0 & 0 & 0 & 0 & 0 & -1 & -1 \\
1 & 1 & 0 & 1 & 1 & 0 & 1 & 0 & 0 & 0 & 0 & 0 & 0 & 0 & 0 & 0 & 0 & 0 & 1 & 1 & 0 & 0 & 0 & -1 \\
1 & 1 & 0 & 1 & 1 & 0 & 1 & 0 & 0 & 0 & 0 & 0 & 0 & 0 & 0 & 0 & 0 & 0 & -1 & 1 & 0 & 0 & 0 & -1 \\
1 & -1 & -1 & -1 & 1 & -1 & 0 & 0 & 1 & 0 & 0 & 0 & 0 & 1 & 0 & 0 & 0 & 0 & 0 & 0 & 0 & 0 & 0 & 0 \\
1 & -1 & 0 & 0 & 1 & 0 & -1 & 0 & 1 & 1 & 0 & 0 & 0 & 0 & 0 & 1 & 0 & 0 & 1 & 0 & 0 & 0 & 0 & 0 \\
1 & -1 & 0 & 0 & 1 & 0 & -1 & 0 & 1 & 1 & 0 & 0 & 0 & 0 & 0 & -1 & 0 & 0 & 1 & 0 & 0 & 0 & 0 & 0 \\
1 & 0 & 1 & 0 & 1 & 0 & -1 & 0 & 0 & 0 & -1 & 1 & 0 & 1 & 0 & 0 & 0 & 0 & 0 & 0 & 0 & 0 & -1 & 0).
\end{equation}
Magma confirms that the sublattice is isomorphic to $\sqrt{2} \Lambda_L$. Analogous data giving superconformal sublattices of the 23 remaining Niemeier lattices are listed in a supplementary text file attached to the arXiv posting for this work.

\clearpage

\bibliographystyle{JHEP}
\bibliography{bib.bib}

\providecommand{\href}[2]{#2}\begingroup\raggedright\begin{thebibliography}{10}

\bibitem{moore_beauty_2023}
G.W.~Moore and R.K.~Singh, \emph{Beauty {And} {The} {Beast} {Part} 2: {Apprehending} {The} {Missing} {Supercurrent}},  Sept., 2023.

\bibitem{dixon_beauty_1988}
L.~Dixon, P.~Ginsparg and J.~Harvey, \emph{Beauty and the beast: {Superconformal} symmetry in a monster module}, \href{https://doi.org/10.1007/BF01217740}{\emph{Communications in Mathematical Physics} {\bfseries 119} (1988) 221}.

\bibitem{frenkel_vertex_1988}
I.B.~Frenkel, J.~Lepowsky and A.~Meurman, \emph{Vertex operator algebras and the {Monster}}, no.~v. 134 in Pure and applied mathematics, Academic Press, Boston (1988).

\bibitem{schellekens_meromorphic_1993}
A.N.~Schellekens, \emph{Meromorphic c=24 {Conformal} {Field} {Theories}}, \href{https://doi.org/10.1007/BF02099044}{\emph{Communications in Mathematical Physics} {\bfseries 153} (1993) 159}.

\bibitem{Hohn:2017dsm}
G.~H{\"o}hn, \emph{{On the Genus of the Moonshine Module}},  8, 2017.

\bibitem{MR4513145}
S.~M\"oller and N.R.~Scheithauer, \emph{Dimension formulae and generalised deep holes of the {L}eech lattice vertex operator algebra}, \href{https://doi.org/10.4007/annals.2023.197.1.4}{\emph{Ann. of Math. (2)} {\bfseries 197} (2023) 221}.

\bibitem{MR4200469}
J.~van Ekeren, C.H.~Lam, S.~M\"oller and H.~Shimakura, \emph{Schellekens' list and the very strange formula}, \href{https://doi.org/10.1016/j.aim.2021.107567}{\emph{Adv. Math.} {\bfseries 380} (2021) Paper No. 107567, 33}.

\bibitem{Hohn:2023auw}
G.~H{\"o}hn and S.~M{\"o}ller, \emph{{Classification of Self-Dual Vertex Operator Superalgebras of Central Charge at Most 24}},  \href{https://arxiv.org/abs/2303.17190}{{\ttfamily 2303.17190}}.

\bibitem{MR4810075}
S.~M\"oller and N.R.~Scheithauer, \emph{A geometric classification of the holomorphic vertex operator algebras of central charge 24}, \href{https://doi.org/10.2140/ant.2024.18.1891}{\emph{Algebra Number Theory} {\bfseries 18} (2024) 1891}.

\bibitem{duncan_super-moonshine_2006}
J.F.~Duncan, \emph{Super-moonshine for {Conway}'s largest sporadic group},  Sept., 2006.
\newblock 10.48550/arXiv.math/0502267.

\bibitem{kapustin_fermionic_2017}
A.~Kapustin and R.~Thorngren, \emph{Fermionic {SPT} phases in higher dimensions and bosonization}, \href{https://doi.org/10.1007/JHEP10(2017)080}{\emph{Journal of High Energy Physics} {\bfseries 2017} (2017) 80}.

\bibitem{Gaiotto:2018ypj}
D.~Gaiotto and T.~Johnson-Freyd, \emph{{Holomorphic SCFTs with small index}}, \href{https://doi.org/10.4153/S0008414X2100002X}{\emph{Can. J. Math.} {\bfseries 74} (2022) 573} [\href{https://arxiv.org/abs/1811.00589}{{\ttfamily 1811.00589}}].

\bibitem{benjamin_extremal_2015}
N.~Benjamin, E.~Dyer, A.L.~Fitzpatrick and S.~Kachru, \emph{An {Extremal} {N}=2 {Superconformal} {Field} {Theory}}, \href{https://doi.org/10.1088/1751-8113/48/49/495401}{\emph{Journal of Physics A: Mathematical and Theoretical} {\bfseries 48} (2015) 495401}.

\bibitem{harrison_extremal_2016}
S.M.~Harrison, \emph{Extremal chiral $\mathcal{N}=4$ {SCFT} with $c=24$}, \href{https://doi.org/10.1007/JHEP11(2016)006}{\emph{Journal of High Energy Physics} {\bfseries 2016} (2016) 6}.

\bibitem{witten_three-dimensional_2007}
E.~Witten, \emph{Three-{Dimensional} {Gravity} {Revisited}},  June, 2007.
\newblock 10.48550/arXiv.0706.3359.

\bibitem{dolan_conformal_1990}
L.~Dolan, P.~Goddard and P.~Montague, \emph{Conformal field theory of twisted vertex operators}, \href{https://doi.org/10.1016/0550-3213(90)90644-S}{\emph{Nuclear Physics B} {\bfseries 338} (1990) 529}.

\bibitem{dolan_conformal_1996}
L.~Dolan, P.~Goddard and P.~Montague, \emph{Conformal {Field} {Theories}, {Representations} and {Lattice} {Constructions}}, \href{https://doi.org/10.1007/BF02103716}{\emph{Communications in Mathematical Physics} {\bfseries 179} (1996) 61}.

\bibitem{Dolan:1989kf}
L.~Dolan, P.~Goddard and P.~Montague, \emph{{Conformal Field Theory, Triality and the Monster Group}}, \href{https://doi.org/10.1016/0370-2693(90)90821-M}{\emph{Phys. Lett. B} {\bfseries 236} (1990) 165}.

\bibitem{dong_associative_1996}
C.~Dong, H.~Li, G.~Mason and S.P.~Norton, \emph{Associative subalgebras of the {Griess} algebra and related topics},  July, 1996.
\newblock 10.48550/arXiv.q-alg/9607013.

\bibitem{hohn_systematic_2022}
G.~Höhn and S.~Möller, \emph{Systematic {Orbifold} {Constructions} of {Schellekens}' {Vertex} {Operator} {Algebras} from {Niemeier} {Lattices}}, \href{https://doi.org/10.1112/jlms.12659}{\emph{Journal of the London Mathematical Society} {\bfseries 106} (2022) 3162}.

\bibitem{nebe_catalogue_nodate}
G.~Nebe and N.~Sloane, ``Catalogue of {Lattices}.'' \url{https://www.math.rwth-aachen.de/~Gabriele.Nebe/LATTICES/index.html}.

\end{thebibliography}\endgroup

\end{document}